\documentstyle[aps]{revtex}

\begin{document}
\title{The orbital moment in NiO}
\author{R.J. Radwa\'{n}ski}
\address{Center for Solid State Physics, \'{s}w. Filip 5, 31-150 Krak\'{o}w.\\
Inst. of Physics, Pedagogical University, 30-084 Krak\'{o}w. }
\author{Z. Ropka}
\address{Center for Solid State Physics, \'{s}w. Filip 5, 31-150 Krak\'{o}w, Poland.}
\author{R. Michalski}
\address{Center for Solid State Physics, \'{s}w. Filip 5, 31-150 Krak\'{o}w.\\
Inst. of Physics, Pedagogical University, 30-084 Krak\'{o}w.\\
email: sfradwan@cyf-kr.edu.pl}
\maketitle

\begin{abstract}
The orbital and spin moment of the Ni$^{2+}$ ion in NiO has been calculated
within the quasi-atomic approach. The orbital moment of 0.54 $\mu _B$
amounts at 0 K, in the magnetically-ordered state, to more than 20\% of the
total moment (2.53 $\mu _B$). For this outcome, being in nice agreement with
the recent experimental finding taking into account the spin-orbit coupling
is indispensable.
\end{abstract}

\date{(8.06.1999)}

NiO attracts large attention of the magnetic community by more than 50
years. Despite of its simplicity (two atoms, NaCl structure, well-defined
antiferromagnetism (AF) with T$_N$ of 525 K) and enormous theoretical and
experimental works the consistent description of its properties, reconciling
its insulating state with the unfilled 3d band is still not reached [1-4].

The aim of this short Letter is to report the calculations of the magnetic
moment of NiO. We attribute this moment to the Ni$^{2+}$ ions. We have
calculated the moment of the Ni$^{2+}$ ion in the NiO$_6$ octahedral
complex, its spin and orbital parts, and the orbital moment as large as 0.54 
$\mu _B$ at 0 K has been revealed. The approach used can be called the
quasi-atomic approach as the starting point for the description of a solid
is consideration of the atomic-like structure of the constituting
atoms/ions, in the present case of the Ni$^{2+}$ ions.

We have treated the 8 outer electrons of the Ni$^{2+}$ ion as forming the
highly-correlated electron system 3d$^8$. Its ground term is described by
two Hund's rules yielding S=1 and L=3 i.e. the ground term $^3$F [5]. Such
the localized highly-correlated electron system interacts in a solid with
the charge and spin surroundings. The charge surrounding has the octahedral
symmetry owing to the NaCl-type of structure of NiO. Effect of small
trigonal distortion experimentally observed will be discussed elsewhere. It
turns out that the trigonal distortion is important for the detailed
formation of the AF structure but it only slightly influences the spin and
orbital moments. Our Hamiltonian for NiO consists of two terms: the
single-ion-like term H$_d$ of the 3d$^8$ system and the d-d intersite
spin-dependent term. Calculations somehow resemble those performed for
rare-earth systems, see e.g. Ref. 6. For the calculations of the
quasi-atomic single-ion-like Hamiltonian of the 3d$^8$ system we take into
account the crystal-field interactions of the octahedral symmetry and the
spin-orbit coupling (octahedral CEF parameter B$_4$=+2 meV, the spin-orbit
coupling $\lambda $=-41 meV). The single-ion states under the octahedral
crystal field and the spin-orbit coupling (the NiO$_6$ complex) have been
calculated by consideration of the Hamiltonian: 
\begin{equation}
H_d=B_4(O_4^0+5O_4^4)+\lambda _{s-o}L\cdot S
\end{equation}

These calculations have revealed [7] the existence of the fine electronic
structure with the charge-formed ground state containing three localized
states, originating from the cubic subterm $^3$A$_{2g}$, characterized by
the total moment of 0 and $\pm $2.26 $\mu _B$. For the doublet the orbital
moment amounts to 0.27 $\mu _B$. It, however, fully cancels in the
paramagnetic state and reveals itself only in the presence of the magnetic
field, external or internal in case of the magnetically-ordered state, that
polarizes two doublet states. The intersite spin-dependent interactions
cause the (antiferro-)magnetic ordering. They have been considered in the
mean-field approximation with the molecular-field coefficient n acting
between magnetic moments $m$=(L+2.0023$\cdot $S) $\mu _B$. The value of $n$
in the Hamiltonian

\begin{equation}
H_{d-d}=n\left( -m_i\cdot m_i+\frac 12\left\langle m_i^2\right\rangle \right)
\end{equation}

has been adjusted in order to reproduce the experimentally-observed Neel
temperature. The fitted value of n has been found to be -200T/ $\mu _B$. It
means that the Ni ion in the magnetic state experiences the molecular field
of 510 T (at 0 K).

The calculated value of the magnetic moment at 0 K in the
magnetically-ordered state amounts to 2.53 $\mu _B$. It is built up from the
spin moment of 1.99 $\mu _B$ (S=0.995) and the orbital moment of 0.54 $\mu
_B $. The increase of m$_L$ in comparison to the paramagnetic state is
caused by the further polarization of the ground-state eigenfunction by the
magnetic field. The orbital moment is quite substantial being more than 20\%
of the total moment. Our theoretical outcome, revealing the substantial
orbital moment is in nice agreement with the very recent experimental result
of 2.2$\pm $0.3 $\mu _B$ for the Ni moment at 300 K [8]. This magnetic x-ray
experiment has revealed the orbital moment of 0.32$\pm $0.05 $\mu _B$ and
the spin moment of 1.90$\pm $0.20 $\mu _B$ at 300 K. From the calculated
temperature dependence of the total moment, shown in Fig. 1, one sees that
the calculated moment at 300 K amounts to 2.2 $\mu _B$ fully reproducing the
experimental results.

We would like to point out that the evaluation of the orbital moment is
possible provided the spin-orbit coupling is taken into account. It confirms
the importance of the spin-orbit coupling for the description of the 3d-ion
compounds [7]. The present model allows, apart of the ordered moment and its
spin and orbital components to calculate many physically important
properties like temperature dependence of the magnetic susceptibility,
temperature dependence of the heat capacity, the spectroscopic g factor, the
fine electronic structure in the energy window below 3 eV with at least 20
localized states. Our calculations indicate that in NiO at 0 K there is set
up the molecular field of 510 T. We have got that the magnetically-ordered
state of NiO has lower energy than the paramagnetic one by 3.25 kJ/mol (=
33.5 meV/ion) at 0 K. Of course, these energies are equal at T$_N$.

In conclusion, the orbital and spin moment of the Ni$^{2+}$ ion in NiO has
been calculated within the quasi-atomic approach. The orbital moment of 0.54 
$\mu _B$ amounts at 0 K, in the magnetically-ordered state, to more than
20\% of the total moment (2.53 $\mu _B$). For this theoretical outcome,
being in nice agreement with the recent experimental finding, taking into
account the intra-atomic spin-orbit coupling is indispensable.

A note added during the referee process (11.01.2000). This Letter is a
subject of the long referee procedure in Phys.Rev.Lett. (Receipt date: 8
June 1999. The validity of this model to NiO is a subject of the scientific
bet with Dr J.Sandweiss, the Div. Ass. Editor of Phys.Rev.Lett. for 1
million USA\ dollars.

{\bf Fig. 1.} Temperature dependence of the Ni$^{2+}$-ion moment in NiO. At
0 K the total moment of 2.53 $\mu _B$ is built up from the orbital and spin
moment of 0.54 and 1.99 $\mu _B$. The calculations have been performed for
the quasi-atomic parameters of the octahedral crystal field B$_4$= +2 meV,
the spin-orbit coupling constant $\lambda _{s-o}$= -41 meV and intersite
spin-dependent interactions given by the molecular-field coefficient n =
-200 T/ $\mu _B$.

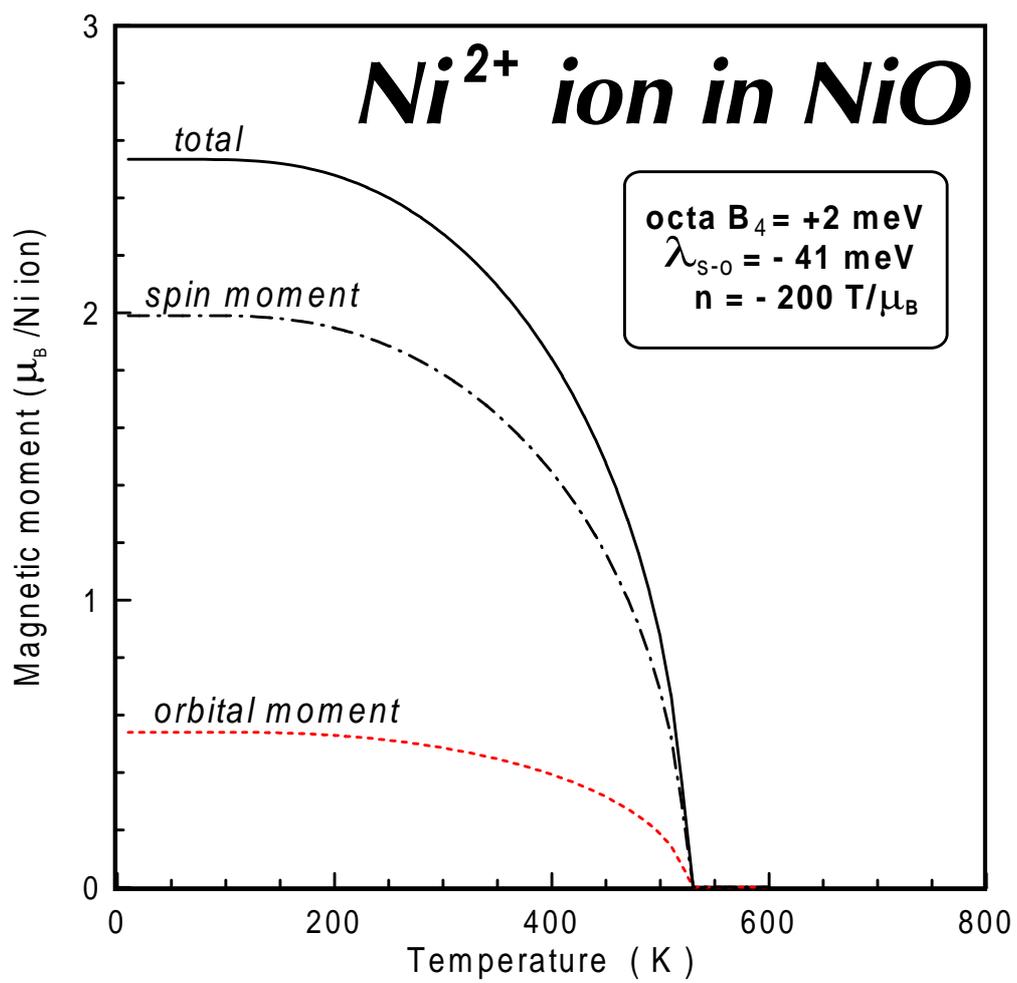
\end{document}